# Seasonal prediction of climate-driven fire risk for decision-making and operational applications in a Mediterranean region

Marco Turco[1,2*], Raül Marcos-Matamoros[1,2], Xavier Castro[3], Esteve Canyameras[3], Maria Carmen Llasat[1]

[1]Department of Applied Physics, University of Barcelona, Av. Diagonal 647, Barcelona 08028, Spain
[2]Barcelona Supercomputing Center – Centro Nacional de Supercomputación (BSC-CNS), c/ Jordi Girona 29, Barcelona 08034, Spain
[3]SPIF (Forest Fire Prevention Office - Generalitat of Catalonia), Barcelona 08028, Spain

[*] Corresponding author: turco.mrc@gmail.com

**Abstract**

In this paper, we assess and develop a climate service focused on the production of seasonal predictions for summer wildfires in a Mediterranean region through a participatory approach with end-users. We start by building a data-driven model that links a drought indicator (Standardised Precipitation Evapotranspiration Index; SPEI) with a series of burned areas in Catalonia (northeastern Spain). Afterwards, we feed this model with SPEI forecasts obtained through a combination of the antecedent observed conditions and climatology. Finally, we assess the forecasting skill of the system by using cross-validation to evaluate the predictions as if they had been made operationally. Our fire forecasting system reveals an untapped and useful burned area predictive ability. We argue that this source of predictability is mostly attributable to the effect of observed initial conditions on summer drought conditions. This system was conceived with the stakeholders, merging climate-driven predictions with information that is of interests to the users, including the identification of climate variables, thresholds and models. The co-production of this customized system allows fire-risk outlooks to be translated into usable information for fire management. This fire forecasting ability plays a crucial role in developing proactive fire management practices such as long-term fuel assessment and other fire-risk planning, thus minimising the impact of adverse climate conditions on summer burned area.

**Highlights**

- A climate service for seasonal predictions for summer wildfires is proposed
- Logistic regression models linking drought and fire variables are defined
- Persistence-based drought forecasts feed this data-driven fire model
- Summer fires can be forecast months in advance with this strategy
- The co-production of this system is key to providing usable information

**Introduction**

The Mediterranean is a high fire-risk region, where fires are the cause of severe economic and environmental losses and even human casualties (Keeley et al., 2011; Moreira et al., 2011; San-Miguel-Ayanz et al., 2013). In recent decades, the annual burned area and number of fires have generally decreased in this region (Turco et al., 2013a, 2016). The increased efforts in fire suppression have probably played an important role in driving the general downward trends described for most of the Mediterranean area (Moreno et al., 2014; Ruffault and Mouillot 2015). Indeed, in recent decades fire management strategies have been improved thanks to new technologies and experience, while climate drivers would probably have led to an opposite trend (see, e.g., Amatulli et al., 2013; Bedia et al., 2013; Batlori et al, 2013; Turco et al., 2014; Jolly et al., 2015; Dupire et al., 2017; Fréjaville and Curt 2017). Nevertheless, the increase in societal exposure to large wildfires in recent years (Bowman et al., 2017), along with the effects of climate change, might still overcome current fire prevention efforts and thus more and different fire management approaches must be considered in order to increase our resilience towards future Mediterranean forest fires (Moritz et al., 2014). In other words, keeping fire management actions at the current level might not be enough to compensate for a projected future increase in droughts and high temperatures (Turco et al., 2018a).

While today fire management strategies in the Mediterranean Europe focus mainly on fire exclusion and suppression (Mavsar et al., 2013), the efficacy of these long-term approaches could be rather limited, as they can lead to higher fuel load and fuel connectivity (Miller and Urban 2000; Collins et al., 2013). Future fire management should therefore expand prevention and adaptation measures, in addition to suppression (Duguy et al., 2013; Fernandes et al., 2013; Alcasena et al., 2018).

Climate variability has a significant influence on fires (Pausas, 2004; Pereira et al., 2005; Aldersley et al., 2011; Koutsias et al., 2013; Amraoui et al., 2015; Urbieta et al., 2015; Bedia et al., 2015; Abatzoglou and Williams, 2016; Paschalidou and Kassomenos, 2016; Pérez-Sánchez et al., 2017; Turco et al., 2013b, 2017a,b). Indeed, the dependence of wildfires on weather and climatic conditions means that fire risk has a certain level of predictability at several time scales (see, e.g., Chen et al., 2016; Di Giuseppe et al., 2016; Turco et al., 2018b). Hence, early prediction of unfavourable conditions can help to mitigate adverse fire effects by supporting the decision-making by policymakers and civil protection agencies with regard to fuel management and resource allocation (Mavsar et al., 2013; San-Miguel-Ayanz et al., 2013). For instance, the ability to forecast the risk of fire months in advance makes it possible to optimize fire-fighting resources (e.g., fire-fighters, equipment and aircraft) and target specific burning restrictions. Also, seasonal forecasts may help to increase preparedness in vulnerable areas such as wildland-urban interfaces and support decisions to reduce fuel load and continuity with prescribed burning and fuel breaks (Moreno et al., 2014; Alcasena et al., 2018). Thus, developing weather-induced fire alarms is key in developing adequate fire management strategies.

Seasonal climate forecasts represent an important tool to inform end-users and are increasingly used across a range of application areas (Lemos et al., 2012; Iizumi et al., 2013; Doblas-Reyes et al., 2013; White et al., 2017; Ceglar et al., 2018; Soares et al., 2018). However, even though additional value for fire applications in several regions of the world has been shown (Chen et al., 2011; Fernandes et al., 2011; Gudmundsson et al., 2014; Marcos et al., 2015; Spessa et al., 2015; Shawki et al., 2017; Turco et al., 2018b), seasonal climate forecasts of fire risk have faced numerous challenges to adequately meet end-users' expectations. This is mainly due to limitations in observations, difficulties in disentangling the many determinants of fire and in translating climate-fire predictions into useful information. Indeed, to the best of our knowledge, there is no operational activity available that provides fire-risk predictions to end-users in Mediterranean regions. For instance, the European Forest Fire Information System (EFFIS) established by the Joint Research Centre and the Directorate General for the Environment of the European Commission provides seasonal forecast maps showing the temperature and rainfall anomalies expected to prevail over European and Mediterranean areas in the following 2 months. These maps are based on the ECMWF (European Centre for Medium-Range Weather Forecasts) Seasonal Forecasting System. The maps show the regions that are expected to be warmer/colder and dryer/wetter (than normal) in Europe and Mediterranean countries (http://effis.jrc.ec.europa.eu/applications/seasonal-forecast/), but a direct estimation of the expected fire anomalies is missing. Probably, the low skill of dynamical forecast systems in variables such as precipitation in key regions like Europe (Frías et al., 2010; Doblas-Reyes et al., 2013; Turco et al., 2017c; Bedia et al., 2018) and the lack of communication/training between seasonal forecast producers and end-users have restricted the applicability of these long-term forecasts (Soares and Dessai, 2016).

In this contribution, we assess the skill of climate predictions of burned area and whether they meet the expectations and requests from the end-users. Specifically, we present a pilot climate service developed for the Forest Fire Prevention Office of the Generalitat de Catalunya (SPIF). Here we show that a seasonal climate forecast of fire risk, when provided through a service tailored to the user's needs, can enable a more effective adaptation to climate variability and change, offering an under-exploited opportunity to minimise the impacts of adverse climate conditions. The engagement of the users starts with the identification of climate variables, thresholds and events relevant to those users. These elements are then analysed to determine whether they can be skilfully predicted at appropriate time scales and whether they meet the expectations and requests from the end-users. We will show that our study has led to a number of original results and methodological approaches that might contribute to the application of seasonal forecasting for fire management in the form of climate services, illustrating the relative merits of climate forecast information to users and the cornerstones of climate stories that engage the users in the co-production of climate information.

The paper is organized as follows. In the next section, the general architecture of the system is introduced. In the "Results" section, a performance analysis of the overall system is presented. Finally, some concluding remarks are made.

**Methods**

Fire and drought data

Accurate data for burned area during the period 1983–2017 were obtained from the Forest Fire Prevention Office of the 'Generalitat de Catalunya' (SPIF). Although there are also fire records for previous years, the database can be considered homogeneous only after 1983 since prior to this date there are no records for one Catalan province (Lleida). Another aspect that may affect the homogeneity of the data is the minimum burned area for which a fire is recorded (see, e.g., Turco et al., 2013, 2016). Indeed, this minimum area is not constant over time: for example, the first part of the database has no records for area smaller than 0.01 ha. To obtain a homogeneous series it is thus necessary to retain only those fires whose area is above a fixed threshold. In the following, we have restricted the analysis to fires with burned area of at least 0.5 ha (however, please note that the total burned area is almost unaffected by this threshold, since it is largely determined by a few large fires: the difference between the total burned area considering all fires and that of fires with burned area larger than 0.5 ha is less than 0.5%). The resulting fire series were finally aggregated to time series of 35 values representing the total burned area of the summer months of July and August in all Catalonia.

We use the standard precipitation and evaporation index (SPEI; Vicente-Serrano et al., 2010) to estimate drought intensity. The SPEI is obtained by a standardizing the multi-month (e.g., 3, 6 or 12 months) water balance values estimated as precipitation minus potential evapotranspiration (PET). In this study, PET is computed using the Thornthwaite equation (Thorntwaite, 1948). Positive values of the SPEI indicate wet conditions relative to the long-term climatology, whereas negative values identify dry conditions. The standardization step is based on a nonparametric approach in which the probability distributions of the data samples are empirically estimated (Hao et al., 2014; Farahmand and AghaKouchak, 2015).

To calculate the observed SPEI, we used two long-term climate databases of observations that provide data in near-real-time, i.e., with updates during the first days of the following month: GHCN-CAMS for two-meter air temperature (Fan Y and Van den Dool, 2008), available for the period starting in 1948 to the near-present with a resolution of 0.5°; and CHIRPS for precipitation data (Funk et al., 2015), available for the period starting in 1981 to the near-present with a spatial resolution of 0.05°. To ensure consistency between the spatial resolutions of these datasets, the CHIRPS dataset was remapped from its original resolution onto the coarsest grid, defined by GHCN-CAMS (0.5° × 0.5°), with a first-order conservative remapping procedure (using Climate Data Operators; https://code.mpimet.mpg.de/projects/cdo). The SPEI was then calculated for the common period 1981-2017 at each point on the 0.5 grid and spatially averaged over Catalonia. PET was calculated using the R package SPEI, version 1.7.

Climate-fire model

This critical step embraces the construction and verification of the empirical model that issues the predictions of the climate service and, therefore, it had to be validated in a way that both demonstrates its predictive skill and is easy for the end-users to understand.

Recent studies have shown that the burned area (BA) by summer fires is directly associated with drought conditions in most sub-regions of Mediterranean Europe (see, e.g., Turco et al., 2017a and 2018a). The approach discussed here builds upon these results and explores the relationship between drought indicators and fires through a statistical model. To develop the empirical (data-driven) climate-fire model we consider the standardized precipitation evapotranspiration index (SPEI). The SPEI is a meteorological drought index that is able to effectively represent the multi-scalar aspect of droughts, and it has been linked to BA variability in many Mediterranean regions (Turco et al., 2017a). In addition, not only is it predictable months in advance (Turco et al., 2017c), but also its links with BA have also been shown to provide successful predictions at seasonal time-scales (Marcos et al., 2015, Turco et al., 2018b). The process of building the empirical model with the SPEI index thus follows these steps (Turco et al., 2017a). First, we normalize the positively skewed BA variables by applying a log transformation (i.e., $Y = \log(BA)$) and the time series of $\log(BA)$ and drought indices are linearly detrended to minimise the influence of slowly-changing factors. Then, to identify the best time window of the SPEI indicator, we (i) compute the correlation between $\log(BA)$ and $SPEI_{sc,8-m}$, with $sc = (6,12)$, and where $8-m$ is the calendar month for which the SPEI is computed, with $m = (0,3)$, i.e., summer and previous spring months; (ii) calculate the significance of the individual correlations (subject to the relationship between BA and SPEI being negative, i.e., a one-tailed hypothesis test); and (iii) owing to the large number of repetitions in the correlation tests (considering different accumulation periods and time-steps), we can expect several correlations to appear significant just by chance, i.e. even if BA is independent of droughts. We address the problem of multiple comparisons with a False Discovery Rate (FDR) test (Ventura et al., 2004). We apply the test on the p-values of the correlations and conservatively set a false rejection rate of $q = 0.05$; and (iv) we seek the minimum correlation values among all the significant correlations calculated in the previous steps. Finally, we fit logistic regressions, as done for instance by Chu et al., (2002) and by Gudmundsson et al., (2014) for fire-risk prediction at seasonal scale. We fit two models, for burned area exceeding the 50th percentile (BA50), and for burned area exceeding the 66th percentile (BA66). These thresholds have been defined by the SPIF as important for fire-risk management in Catalonia, but other regions/users could define other values. The models are of the form:

$$\log(p/(1-p)) = \beta 0 + \beta 1 \cdot SPEI_{sc,(8-m)} + \beta 2 \cdot T + \varepsilon \qquad (1)$$

In Eq.(1), p is the response variable (i.e. the probability of burned area above a certain percentile), *β0* is the intercept; *β1* represents the sensitivity of BA in each region to dry conditions as indicated by the SPEI; *β2* is the coefficient of the time term *T* (in years) that characterises the temporal trends of the fire variable and thus takes into account the possible influence of slowly changing factors over this period of time; and *ε* is a stochastic noise term that captures all other (neglected) processes that influence BA, apart from *SPEI* and *T*. The value of the β coefficients is determined using generalised linear models (GLMs). Drought conditions are measured by the SPEI indices aggregated in multi-month values, *SPEIsc,(8-m)*, where 8-*m* is the calendar month for which the SPEI is computed (e.g., in August when m=0) and *sc* is the time scale (number of months) used to compute the SPEI. These parameters are obtained through the optimization procedure described above.

We tested whether burned area or climate variables present any kind of temporal autocorrelation. Specifically, we firstly detrended (removing a linear trend) all the time series prior to the analysis. This was done because the presence of long-term trends in the time series can bias the correlation analysis. We then computed the autocorrelation of detrended summer burned area values, the detrended precipitation (PRE) and the detrended potential evapotranspiration (PET) series of spring and summer months. This analysis did not produce any significant autocorrelation. Furthermore, neither the spring PET nor the spring PRE exhibited any significant correlation with summer PRE and summer PET, respectively. Thus we can exclude (i) a potential lag effect, where a year with low fire activity is followed by one with above-average fire activity and (ii) that any fake relationship affects the conclusion.

With regard to the operational use of the model, it is important to know whether the forecasts can predict the occurrence of an event where the burned area lies above the median or the 66th percentile. For this reason, it is vital to study the thresholds of probability that activate the alerts in such a way that they maximize the 'hits' (H, the number of times an event is predicted with respect to the number of times it takes place) and minimize the 'false alarms' (F, the relative number of times the event is predicted but does not actually take place). Hence, we considered the relative operating characteristic (ROC) diagram to verify our forecasts. ROC shows the hit rate against the false alarm rate for different potential decision thresholds for different potential decision thresholds. The significance of ROC is estimated using a Mann–Whitney U test (Mason and Graham, 2002). Finally, we used the area under the curve ROC (RA) to show the end-users whether our model is more useful than a random one (when RA > 0.5), considering that the perfect value for this metric is 1.

**Results**

Defining the climate-fire model

In the first place, we use an empirical approach to systematically explore the cross correlations between the detrended SPEI and log(BA) as explained in the Methods section. Figure 1 shows statistically significant negative correlations between BA and the drought index. Interestingly, both SPEI6 and SPEI12 calculated in June, July, or August show significant (i.e., p-value <0.05) correlations with BA. In addition, these correlations are also collectively significant considering the FDR test applied (see section Methods). Since the negative SPEI values correspond to hot and dry conditions, the SPEI negative correlation indicates that the warmer and drier conditions for the same summer lead to a larger BA. Then, the SPEI index showing the highest correlations in absolute value, and which will be used in the logistic model, is the one that is calculated considering the 6-month period water balance between March and August. This result suggests that, overall, spring and summer drought conditions lead to larger burned area values. Specifically, a relatively short period of drought is presumably needed to have a sufficient amount of biomass available for burning (fuel load) in a wildfire. That is, the so-called "Mediterranean scrub", may take weeks or months of arid conditions to dry out.

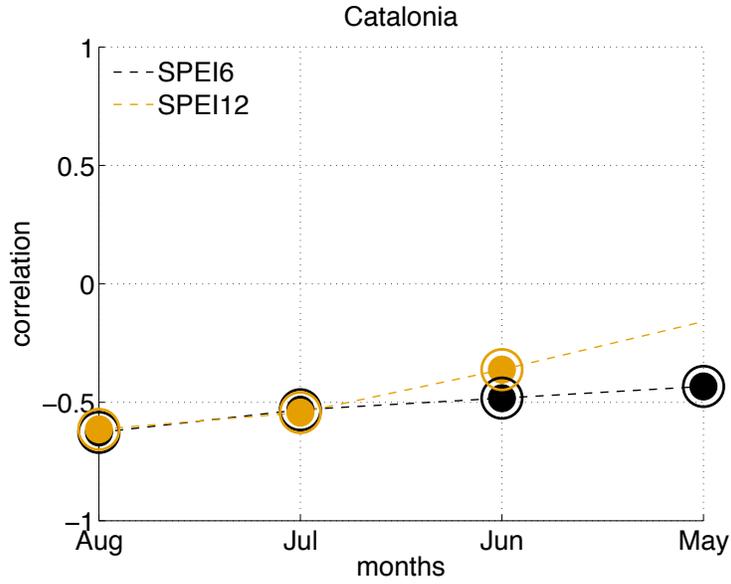

Figure 1. Correlations between the fire variable - detrended log(BA) - and detrended SPEI. The filled circles indicate individual significant correlations (values p <0.05). Open circles around the filled ones indicate correlations that are also collectively significant with an FDR test (Ventura et al., 2004).

Thus the final models for BA50 and BA66 thresholds are:

$$\log(BA50/(1- BA50)) = 0.07 - 1.96 \cdot SPEI_{6,8} - 0.10 \cdot T \qquad (2)$$

$$\log(BA66/(1- BA66)) = -1.20 - 1.50 \cdot SPEI_{6,8} - 1.43\, T \qquad (3)$$

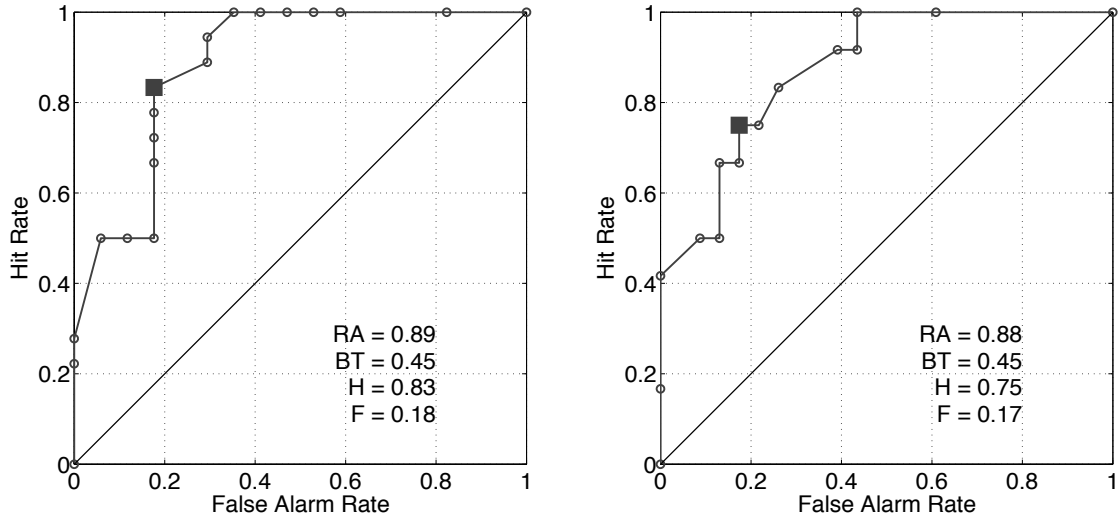

Figure 2. Relative operation characteristics diagram (ROC) for predictions of BA50 (left panel) and BA66 (right panel). The hit rate and false alarm rate values are displayed by the grey open points considering the different threshold probabilities (from 0 to 100% by 5%) of the logistic model (see the running text for more details). The numbers within the plots are the area of ROC (RA), the probability threshold that maximizes the difference between the hit rate and the false alarm rate (BT), the success rate (H) at the BT probability and the false alarm rate (F) at the BT probability.

Figure 2 shows the ROC diagrams for predictions of BA50 and BA66 events. This figure shows that the forecasts have high skill since their curves are above the identity line H=F (when a forecast is indistinguishable from a completely random prediction). The ROC diagram can also help to decide for which forecast value it is better to issue an alarm. Users could decide to take action when the forecast is above the threshold corresponding to a 'very low' fire danger level. For instance, if we set a warning threshold of 0, we would have to issue an alert for every forecast. In this case, all the observed events are forecasted, that is, H=1, but this also implies a high number of false alarms. If, on the other hand, we use a higher threshold, we can reduce the number of false alarms, but at the expense of a greater number of missed events. For instance, if we set a warning threshold of 1, we can never forecast an "event", thus H=0, but neither will we make false alarms.

The choice of the decision threshold is a function both of the skill of the forecast and of the cost/loss ratio of the user. In any case, in a forecasting system affected by uncertainties, missed events can be reduced only by increasing the false alarms, and vice versa. This figure suggests which probability level maximizes the difference between the hit rate and the false alarm rate. This "best threshold" (hereinafter BT) was selected in agreement with the SPIF user, although other values could be considered in other contexts. In the case presented in Figure 2, this threshold is 0.45. This means that if we want to maximise the H – F difference (but please note that users could define other best thresholds according to their cost/loss ratio), an above- 50th percentile burned area event is to be expected when our model predicts a probability higher than 0.45, resulting in H = 0.83 (meaning that 83 out of 100 events are correctly modelled) and F = 0.18 (meaning that 18 out of 100 events were modelled as an "event" which did not actually happen).

These results were obtained considering observed value, while the final objective is to assess the ability to predict BA. The final test is to assess the skill of the BA prediction obtained by feeding the climate-fire model with seasonal forecasts of the SPEI.

As several studies have shown (e.g., Hao et al., 2014, Turco et al., 2017c), purely statistically-based systems have achieved skilful predictions of drought indicators at global or continental scale. Here we adopt a very simple strategy to forecast SPEI values, which consists in merging observational information (for the months previous to the fire season) with climatological values for the upcoming months. For instance, if we forecast the $SPEI_{6,8}$ with the prediction initialized at month 6 (i.e., in June) we calculate the six-month accumulated value of the climatic balance (precipitation minus potential evapotranspiration) in this way:

$$A_{8,2017} = W^{obs}_{3.2017} + W^{obs}_{4.2017} + W^{obs}_{5.2017} + W^{clim}_{6} + W^{clim}_{7} + W^{clim}_{8} \qquad \text{Eq. 4}$$

where $W^{obs}_{i,2017}$ is the observed monthly climatic balance (PRE-PET) for the i-th month of the year 2017 and $W^{clim}_{j}$ is the long-term median (a more robust estimator compared to the mean) of the monthly climatic balance for the j-th month. That is, to calculate the predicted SPEI6 for August, we combine the climatological values of PRE and PET as future forecasts with the antecedent series of observed records. Specifically, Eq. 4 indicates that the first three months (March, April and May) are observed, while the other three (June, July and August) are forecasted using climatological values.

In this study, we assess the skill of forecasts initialized in May, June, July and August. For instance, to predict the $SPEI_{6,8}$ in May, we combined two months (March and April) of observations with four months (May, June, July and August) of forecasts. Importantly, all the forecasts are made by using cross-validation in order to evaluate the predictions as if they had been made operationally. To avoid artificial skill, the logistic regression models were calibrated in each step of the cross-validation. More specifically, a leave-one-out validation method is applied, in which a moving window of one year is used as the validation data and the remaining observations as the training data.

Figure 3 shows that the out-of-sample predictions using the observed SPEI have skill. These results provide the maximum skill of BA using the SPEI-BA model, as they are obtained using the best available climate data (that is, observational references) as drivers. When the SPEI is forecast, the skill increases with the month of initialisation, as expected. That is, the RA values in longer leads (i.e., the 4-month-ahead prediction issue in May) are lower than those of shorter lead forecasts.

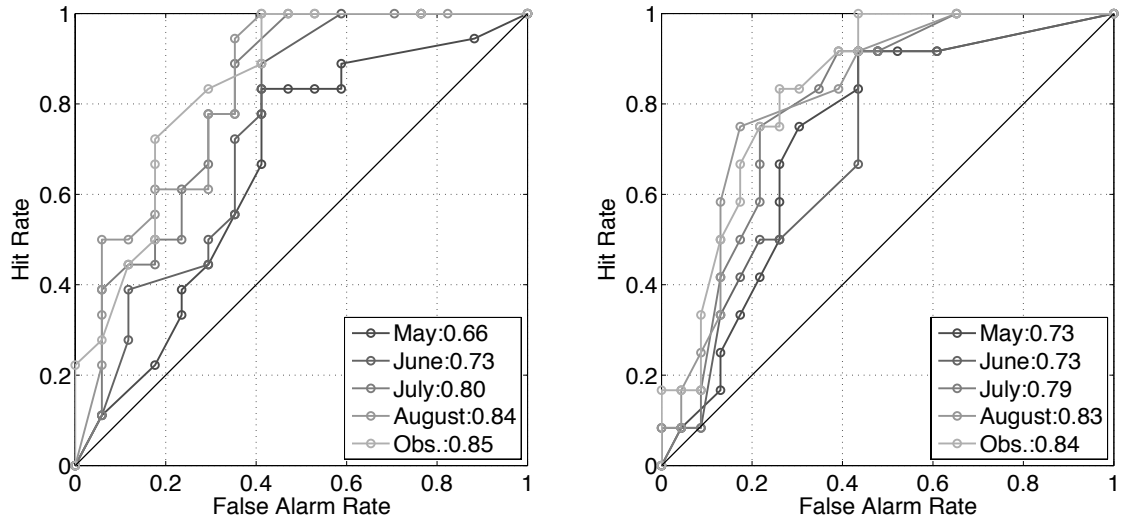

Figure 3. As Figure 2, but for out-of-sample predictions initialized in May, June, July and August. The skill of the out-of-sample predictions with the observed SPEI is also shown for reference. The numbers in the legend indicate the Roc Area of the different forecasts.

Finally, Figure 4 shows the forecast products generated and sent to the SPIF fire managers. Importantly, the users were involved in defining and understanding this plot in order to maximise effectiveness in communicating the prediction and the uncertainties. This figure provides the forecast along with the events observed and past prediction to have a first view of the ability of the skill of the prediction. It includes the forecast (in red), along with the past forecasts and the observation occurrences, thus allowing fast and simple visualization of the past performance of the system. The blue horizontal line highlights the "best thresholds". That is, the level when the hits equal the missed events, the level chosen as the warning threshold in agreement with the end-users, makes efficient decision-making with this tool easy to visualize and communicate. Specifically, the forecasts suggest that the probability of having BA above the 50th percentile is low (around 10%). Before and during the fire season, forecast analyses similar to that shown in Figure 4 are generated at the beginning of each month (i.e., May to August).

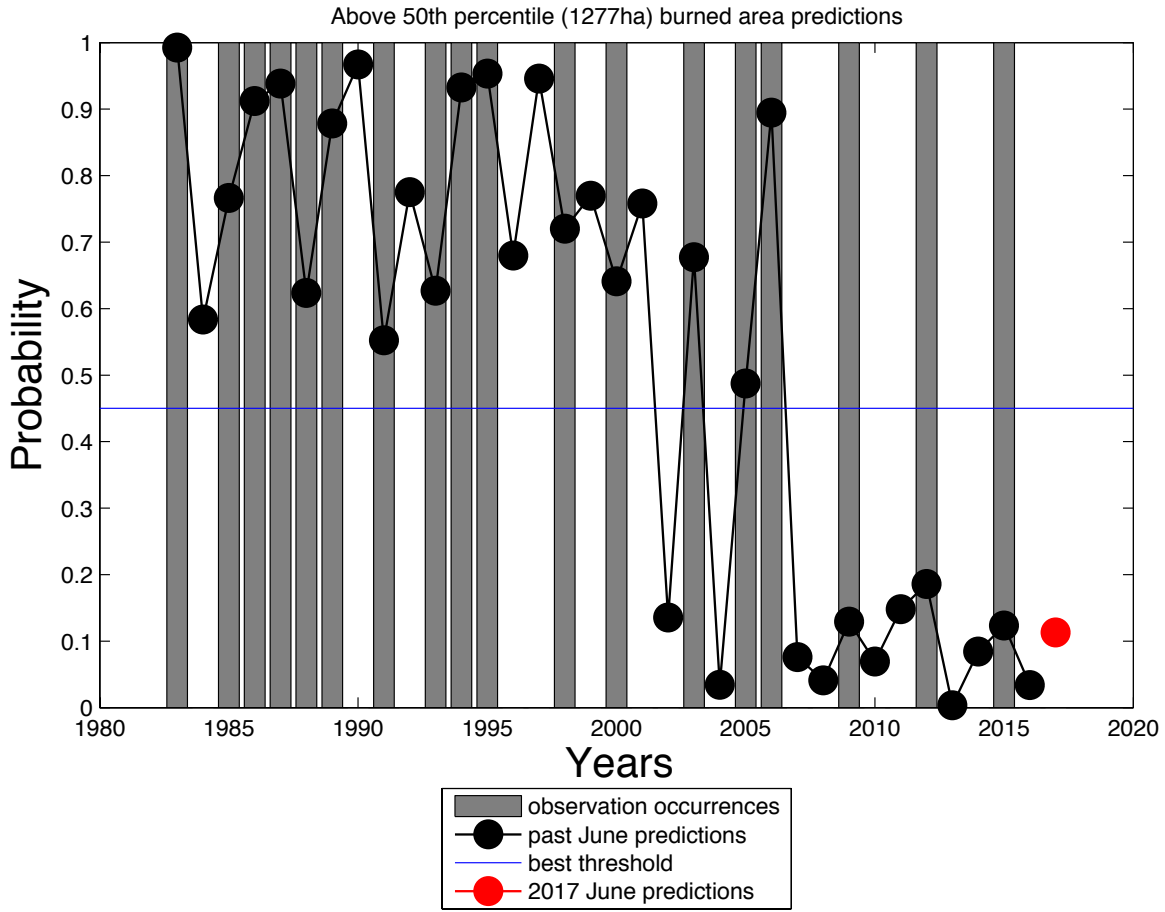

Figure 4. Time series of the probability of BA50. Forecasts issued in June. The red point indicates the 2017 forecasts according to the average conditions.

The verification of the prediction at the end of each fire season is a fundamental step of the producer-user interaction in order to correctly manage expectations. For instance, Figure 4 clearly shows that generally, when a prediction exceeds the BT threshold, an above normal fire season has occurred. However, there are also some missed events, as in 2015, and false alarms, as in 2001. Users should, therefore, be aware of the limitations and uncertainties of the system in order to avoid unrealistic expectations. Finally, as an illustrative example of the information we have shared with end-users, we provide the analysis performed for the 2017 season. Table 1 shows the hectares burned in Catalonia between July and August 2017 and the forecasts for this period: all the forecasts have given values below the percentiles 50 and 66. Given that between July and August 584 hectares were burned, a value lower than the median, we can conclude that the observations were in agreement with the forecasts.

Table 1. Total Burned Area (in hectares) in Catalonia in July and August of 2017. The table shows whether or not the 50th and 66th percentiles was exceeded, together with the corresponding prediction (in green, the correct ones; in red, if any, the errors).

| Observed Burned Area | Prediction | | | | | |
|---|---|---|---|---|---|---|
| | | Value | Prediction to exceed this percentile? | | | |
| | | | M | J | J | A |
| 584 | Percentile 50 | 1301 | NO | NO | NO | NO |
| | Percentile 66 | 3626 | NO | NO | NO | NO |

**Conclusions**

The intensification in drought risks related to climate change further increases both the fire risk in the Mediterranean (Turco et al., 2018a) and the urgency to provide predictive models that can limit its potential fire impacts. In this paper, we have shown that parsimonious drought-fire models are a good tool for developing accurate predictions that can be easily communicated to end-users. This system is intended as a decision aid tool for fire managers, providing an assessment of fire risk in the most critical period of the year (the summer months of July and August). The ultimate goal is to assist fire managers in making proactive decisions to better protect life, human and natural resources, while also increasing fire fighter safety and reduce fire-fighting costs. More specifically, we have described the development and assessment of an operational prototype system that provides information on fire-risk months in advance in a Mediterranean region. In addition, it also presents a novel way to communicate patterns in seasonal fire prediction data. The prototype follows an interdisciplinary approach to develop climate services for the fire-management sector, putting special emphasis on effectively communicating probabilistic predictions to decision-makers. It is the result of a co-production process involving researchers and end-users from the fire management sector.

The system described relies on empirical models fed with near-real time data of climate variables, data publicly available thanks to the Open Data policies of the data provider (see Methods). The use of these data for the application described in this paper helps to promote the so-called "Open Knowledge" approach (http://www.jpi-climate.eu/jpi-strategy/OpenAccessOpenKnowledge). This strategy facilitates knowledge creation, reproduction, transfer and exchange in a more transparent and interactive way, thus providing useful tools to other researchers and, ultimately, contributing to narrowing the gap between research communities and societal actors. In short, open data can create valuable and usable information.

The dialogue with the users during the design and production of the fire prediction system is key to transforming climate information into climate knowledge (Buontempo et al., 2017; Christel et al., 2017). Climate services aim to provide tailored knowledge to stakeholders through research that could be "actionable" (Buizer et al., 2016; van Den Hurk et al., 2016), that is to say, making climate information action-oriented. This study highlights the value of services through an example of their application in the fire-risk sector. However, users need to be well-trained in how to best interpret and use these forecasts, given the actual performance of the seasonal predictions. Until now, the low skill for variables such as precipitation in key regions like Europe has limited the applicability of seasonal climate predictions in these areas. To increase the role of these long-term forecasts in an end-user environment, we have shown their value by using empirical models (data driven) with near-real time climate data. Indeed, the skill found relies largely on merging observational information (for the months previous to the fire season) with empirical seasonal forecasts (for the fire season). This strategy substantially contributes to increase fire predictability, making the most of the best information available to the users.

To sum up, the results suggest that seasonal climate predictions, which are complementary to climate projections, can enable an effective adaptation to both climate variability and change, offering an under-exploited opportunity to minimise the impacts of adverse climate conditions from the present day.

**Acknowledgments**

M.Turco has received funding from the European Union's Horizon 2020 research and innovation programme under the Marie Skłodowska-Curie grant agreement No. 740073 (CLIM4CROP project).

**Author contributions**

M.T. conceived the study, designed and carried out the data analysis and wrote the paper. R.M, X.C.,

E.C. and M.C.L. participated in defining the analysis and methodology, contributed to interpreting the results, and to writing the paper.

**Competing interests**

The authors declare no competing interests.

Turco, M., von Hardenberg, J., AghaKouchak, A., Llasat, M. C., Provenzale, A., & Trigo, R. M.. On the key role of droughts in the dynamics of summer fires in Mediterranean Europe. *Scientific reports*, *7*(1), 81 (2017a).

Urbieta, I. R. et al., Fire activity as a function of fire–weather seasonal severity and antecedent climate across spatial scales in southern Europe and Pacific western USA. Environ. Res. Lett. 10, 114013 (2015).

van Den Hurk, B. J., Bouwer, L. M., Buontempo, C., Döscher, R., Ercin, E., Hananel, C., ... & Pappenberger, F. (2016). Improving predictions and management of hydrological extremes through climate services: www. imprex. eu. Climate Services, 1, 6-11.

Ventura, V., Paciorek, C. J., & Risbey, J. S. (2004). Controlling the proportion of falsely rejected hypotheses when conducting multiple tests with climatological data. Journal of Climate, 17(22), 4343-4356.

Vicente-Serrano, S. M., Beguería, S., & López-Moreno, J. I. (2010). A multiscalar drought index sensitive to global warming: the standardized precipitation evapotranspiration index. Journal of climate, 23(7), 1696-1718.

White, C. J., Carlsen, H., Robertson, A. W., Klein, R. J., Lazo, J. K., Kumar, A., ... & Bharwani, S. (2017). Potential applications of subseasonal-to-seasonal (S2S) predictions. Meteorological Applications.
21